\documentclass[aps,preprint,showpacs]{revtex4}
\usepackage{epsfig}
\begin{document}
%%%%%%%%%%%%%%%%%%%%%%%%%%%%%%%%%%%%%%%%%%%%%%%%%%%%%%%%%%
\newcommand{\pr}{^{\prime}}
\newcommand{\bfx}{{\bf x}}
\newcommand{\bfy}{{\bf y}}
\newcommand{\bfz}{{\bf z}}
\newcommand{\bfp}{{\bf p}}
\newcommand{\la}{\langle}
\newcommand{\ra}{\rangle}
\newcommand{\rp}{{p}}
\newcommand{\rpp}{{{p}^{\prime}}}
\newcommand{\rx}{{x}}
\newcommand{\eps}{\varepsilon}
\newcommand{\be}{\begin{eqnarray}}
\newcommand{\ee}{\end{eqnarray}}
\newcommand{\ba}{\begin{array}}
\newcommand{\ea}{\end{array}}
\newcommand{\balpha}{{\mbox{\boldmath$\alpha$}}}
%%%%%%%%%%%%%%%%%%%%%%%%%%%%%%%%%%%%%%%%%%%%%%%%%%%%%%%%%%
\title{Evaluation of the two-photon exchange diagrams for the
$\bf (1s)^22p_{3/2}$ electron configuration in Li-like ions}
\author{A.~N.~Artemyev$^{1,2,3}$, V.~M.~Shabaev$^{1,2,3}$, 
M.~M.~Sysak$^{1,3}$,
 V.~A.~Yerokhin$^{1}$, T.~Beier$^3$, G.~Plunien$^2$,
and G.~Soff$^2$}
\affiliation{$^1$Department of Physics, St. Petersburg State University,
Oulianovskaya 1, Petrodvorets, St. Petersburg 198504, Russia\\
$^2$Institut f\"{u}r Theoretische Physik, TU Dresden, Mommsenstra{\ss}e
13, D-01062 Dresden, Germany\\
$^3$Gesellschaft f\"{u}r Schwerionenforschung, Planckstrasse 1,
D-64291 Darmstadt, Germany
}
\date{\today}
\begin{abstract}
We present {\it ab initio} calculations of the complete gauge-invariant set
of two-photon exchange graphs for the $(1s)^22p_{3/2}$ electron configuration
in Li-like ions. These calculations are an important step towards the precise
theoretical determination of the $2p_{3/2}$-$2s$ transition energy in the
framework of QED.
\end{abstract}
\maketitle

\section{Introduction}

At present the lowest-lying states in heavy Li-like ions can be investigated 
very precisely both theoretically and experimentally. 
One of the most precise experimental
results in these systems has been obtained by Beiersdorfer and co-workers
\cite{beiersdorfer98} for the $2p_{3/2}$-$2s$ transition energy in Li-like
bismuth, which was determined with an accuracy of $0.04$ eV. Accurate
experimental data are available at present for a number of other elements as
well. For the latest high-precision measurements we refer  to Refs.
\cite{bosselmann99,feili00,brandau00}; the outline of earlier investigations
can be found in Ref. \cite{bosselmann99}.

The accuracy reached in experimental investigations provides a promising tool
for probing QED corrections in the
strong Coulomb field of the nucleus up to second order in the fine structure
constant $\alpha$. 
For the $2p_{1/2}$-$2s$ transition, this
project has been carried out in a series of our previous investigation
\cite{artemyev99,yerokhin99,yerokhin00,yerokhin01pra2p}. In Ref.
\cite{yerokhin01pra2p} we completed the evaluation of all two-electron QED
corrections of second order in  $\alpha$ and
obtained most accurate theoretical predictions for the $2p_{1/2}$-$ 2s$ splitting
within a wide range of nuclear charge numbers $Z$. Based on a careful estimate
of the uncertainty of the theoretical values, we concluded that already now
the comparison of theory and experiment for Li-like uranium provides a test
of QED effects of second order in $\alpha$  at the level of accuracy
 of about 17\%. For
the $2p_{3/2}$-$2s$ and $2p_{1/2}$-$2s$ transitions in Li-like bismuth,
analogous calculations have been performed recently by Sapirstein and Cheng
\cite{sapirstein01}. However, in order to match with the experimental accuracy for
the $2p_{3/2}$-$2s$ splitting, rigorous evaluations of second-order QED
corrections are required also for other ions than bismuth. The first step in
this direction has been performed in our earlier investigation \cite{artemyev99} where
we have evaluated the vacuum-polarization screening correction for several energy
levels of Li-like ions, including the $(1s)^22p_{3/2}$ state. The aim of the
present work is to calculate the two-photon exchange correction for this
state (for extensive calculations of these corrections for the
lower states in Li-like ions and for non-mixed low-lying states 
in He-like ions we refer the reader to Refs.
\cite{yerokhin00,yerokhin01pra2p,blundell93b,lindgren95pra,moh00,and01,and02}).
 After all that, the self-energy screening correction remains the last
uncalculated two-electron second-order QED contribution for this state.

This paper is organized as follows. In the next section we present the basic
formulas for the two-photon exchange correction for the $(1s)^22p_{3/2}$
state. The description of our numerical procedure is given in Sec. III,
and the results obtained are discussed in Sec. IV.

Relativistic units ($\hbar = c = 1$) are used throughout this paper.

%%%%%%%%%%%%%%%%%%%%%%%%%%%%%%%%%%%%%%%%%%%%%%%%%%%%%%%%%%%%%%%%%%%%%%
%%%%%%%%%%%%%%%%%%%%%%%%%%%%%%%%%%%%%%%%%%%%%%%%%%%%%%%%%%%%%%%%%%%%%%
%%%%%%%%%%%%%%%%%%%%%%%%%%%%%%%%%%%%%%%%%%%%%%%%%%%%%%%%%%%%%%%%%%%%%%
\section{Basic formulas}

The detailed derivation of the two-photon exchange corrections to the
$(1s)^22s$ and $(1s)^22p_{1/2}$ states of Li-like ions can be found in our
previous paper \cite{yerokhin01pra2p}. For the $(1s)^22p_{3/2}$ state the
derivation is performed along the same lines. Thus, we present mainly the
final formulas here. Our derivation is based on the two-time Green function
(TTGF) method \cite{shabaev90ivf, shabaev94ttg2}. For the detailed
description of the method we refer to the recent review \cite{shabaev02rep}.

\begin{figure*}
\epsfig{file=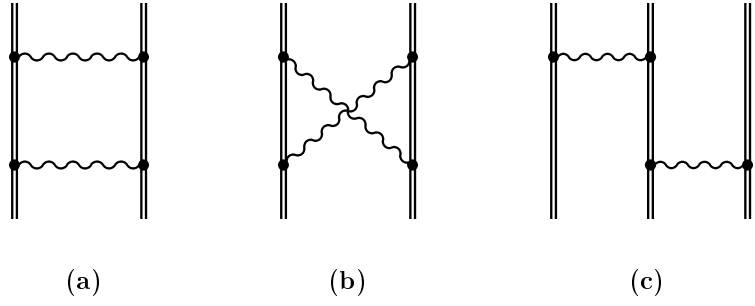,width=0.7\textwidth}
\caption{\label{diagrams}
Feynman diagrams for the two-photon exchange corrections.
}
\end{figure*}

The two-photon exchange corrections to the $(1s)^22p_{3/2}$ state of the
Li-like ions can be conveniently separated in three parts: the two-photon
exchange contribution due to the interaction between two $1s$ electrons, the
two-photon exchange contribution due to the interaction between the valence
electron and one of the $1s$ electrons, and the three-electron contribution. The
first part coincides with the two-photon exchange correction to the
ground-state energy of He-like ions. Its calculation was carried out in
\cite{blundell93b, lindgren95pra}. This correction does not contribute to the
$2p$-$2s$ splitting in Li-like ions and is not considered here. The remaining
two-electron and three-electron corrections are diagrammatically depicted in Fig.
\ref{diagrams}.

We start from the expression for the second-order correction to the energy
shift of the level $k$ \cite{shabaev02rep},
\be\label{secord} \Delta E^{(2)}_k&=&\frac{1}{2\pi
i}\oint_\Gamma dE\, \Delta E\, \Delta g^{(2)}_{kk}(E)\nonumber \\ &-&
\frac{1}{2\pi i}\oint_\Gamma dE\, \Delta E\Delta g^{(1)}_{kk}(E)
\frac{1}{2\pi i}\oint_\Gamma dE\pr\, \Delta g^{(1)}_{kk}(E\pr)\,,
\ee
where $\Delta g_{kk}(E)=g_{kk}(E)-g_{kk}^{(0)}(E)$, $g_{kk}(E)=\la
u_k|g(E)|u_k \ra$, $u_k$ is the unperturbed wave function, 
$\Delta E=E-E_k^{(0)}$, $E_k^{(0)}$ is the unperturbed energy
of the state $k$, and
$g_{kk}^{(0)}(E)=(\Delta E)^{-1}$ is the function $g_{kk}(E)$ in the
zeroth-order approximation. The function $g(E) \equiv g(E,\bfx_1\pr,\cdots
,\bfx_N\pr;\bfx_1,\cdots ,\bfx_N)$ is the temporal Fourier transform 
 of the N-electron two-time Green function.
Its definition and the corresponding Feynman rules can be found in
\cite{shabaev02rep}. The superscripts in Eq.~(\ref{secord}) indicate the
order of the contribution in $\alpha$.

For the two-photon exchange correction, Feynman diagrams contributing to
$\Delta g^{(2)}(E)$ are presented in Fig. \ref{diagrams}. We refer to the
corresponding contributions as the {\it ladder} (a), the {\it crossed} (b),
and the {\it three-electron} (c) terms. The second term in Eq. (\ref{secord})
is known as the disconnected contribution. It vanishes completely  when
considered together with the reducible contribution (for details, see
\cite{yerokhin01pra2p}). In our case, the
unperturbed wave function is
 \be\label{wf}
u_k=\frac{1}{\sqrt{3!}}\sum_P(-1)^P\psi_{Pa}(\bfx_1)\psi_{Pb}(\bfx_2)
\psi_{Pv}(\bfx_3)\,,
\ee
where $v$ denotes the valence electron, $a$ and $b$ are the electrons in
$(1s)^2$ shell, and $P$ is the permutation operator (in the factor
$(-1)^P$, the parity of the permutation is implied by $P$). 
 For brevity we will use
also the following notations:
\be
I(\omega)&=&e^2\alpha^\mu\alpha^\nu D_{\mu\nu}(\omega)\,,\\
I_{abcd}(\omega)&=&\la ab|I(\omega)|cd \ra\,,\\
I_{ab;cd}&=&I_{abcd}(\Delta_{bd})-I_{bacd}(\Delta_{ad})\,,\\
I\pr(\omega)&=&\frac{dI(\omega)}{d\omega}\,,
\ee
where $\Delta_{ab}=\eps_a-\eps_b$, $\alpha^\mu=(1,\balpha)$ are the Dirac
matrices, and $D_{\mu\nu}(\omega)$ is the photon propagator.

We separate the contributions of the diagrams under consideration into two
parts: the {\it reducible}, with the energy of the intermediate state
coinciding with the energy of the initial (final) state, and the {\it irreducible},
for the remainder, respectively. Omitting the derivation similar to that in Ref.
\cite{yerokhin01pra2p}, we present here only the final expressions for the energy shift.
The reducible ("red") and irreducible ("ir") three-electron contributions
read
\be
\Delta E_{\rm ir}^{\rm 3el}&=&\sum_{PQ}(-1)^{P+Q}\nonumber \\
&\times&\sum_{n}\,\pr\frac{I_{P2P3nQ3} (\Delta_{P3Q3})
\,I_{P1nQ1Q2}(\Delta_{Q1P1})}{\eps_{Q1}+\eps_{Q2}-\eps_{P1}
-\eps_n}\,,\,\,\,\,\,\,\,\,\,\,\label{3elirr}\\ \Delta E_{\rm red}^{\rm
3el}&=&\sum_{\mu_a}\biggl[I\pr_{vaav}(\Delta_{va}) (I_{ab;ab}-I_{bv;bv})
\nonumber \\ &+& \frac 12 I\pr_{av\tilde{v}b}(\Delta_{va})
I_{b\tilde{v};av}+\frac12 I\pr_{b\tilde{v}va}(\Delta_{va})
I_{va;\tilde{v}b}\biggr]\,,\label{3elred}
\ee
where $P$ and $Q$ are the permutation operators, and the prime in the sum in
Eq. (\ref{3elirr}) indicates that terms with the vanishing denominator should
be omitted in the summation. In Eq.~(\ref{3elred}) $a$ and $b$ denote $1s$ 
electrons with opposite angular-momentum projections $\mu_a = -\mu_b$, $v$ stands for the
valence state with the angular-momentum projection $\mu_v$, and $\tilde{v}$ is
the valence state with $\mu_{\tilde{v}}=2\mu_a+\mu_v$ (the corresponding 
contribution is assumed to be zero when
$\mu_{\tilde{v}}$ is out of the range $-j_v,\ldots,j_v$).

The irreducible two-electron contribution is
\be \label{2elirr}
\Delta E^{\rm 2el}_{\rm "ir"}&=&\Delta E^{\rm lad}_{\rm dir}+\Delta E^{\rm lad}_{\rm exch}+
\Delta E^{\rm cr}_{\rm dir}+\Delta E^{\rm cr}_{\rm exch}\,,\\
\label{eladdir}
\Delta E^{\rm lad}_{\rm dir}&=&
\sum_{n_1n_2}{}\!\pr
\frac{i}{2\pi}
\int_{-\infty}^{\infty}d\omega
\nonumber\\&\times &
\frac{F^{\rm lad}_{\rm dir}(\omega,n_1n_2)}
{(\eps_c-\omega-\eps_{n_1}u)(\eps_v+\omega-\eps_{n_2}u)}
\,,\\
\label{eladexch}
\Delta E^{\rm lad}_{\rm exch}&=&-\sum_{n_1n_2}{}\!\pr
\frac{i}{2\pi}
\int_{-\infty}^{\infty}d\omega
\nonumber\\&\times &
\frac{F^{\rm lad}_{\rm exch}(\omega,n_1n_2)}
{(\eps_v-\omega-\eps_{n_1}u)(\eps_c+\omega-\eps_{n_2}u)}
\,,\\
\label{ecrdir}
\Delta E^{\rm cr}_{\rm dir}&=&
\sum_{n_1n_2}{}\!\pr
\frac{i}{2\pi}
\int_{-\infty}^{\infty}d\omega
\nonumber\\&\times &
\frac{F^{\rm cr}_{\rm dir}(\omega,n_1n_2)}
{(\eps_c-\omega-\eps_{n_1}u)(\eps_v-\omega-\eps_{n_2}u)}
\,,\\
\label{ecrexch}
\Delta E^{\rm cr}_{\rm exch}&=&-
\sum_{n_1n_2}{}\!\pr
\frac{i}{2\pi}
\int_{-\infty}^{\infty}d\omega
\nonumber\\&\times &
\frac{F^{\rm cr}_{\rm exch}(\omega,n_1n_2)}
{(\eps_v-\omega-\eps_{n_1}u)(\eps_v-\omega-\eps_{n_2}u)}\,.
\ee
Here we introduced the labels "lad" and "cr" for the ladder and the crossed diagram, and
"dir" and "exch" for the direct and the exchange parts. The other notations
are:
\be
F^{\rm lad}_{\rm dir}(\omega,n_1n_2)&=&\sum_{\mu_c\mu_{n_1}\mu_{n_2}}
I_{cvn_1n_2}(\omega)I_{n_1n_2cv}(\omega)\,,\\
F^{\rm lad}_{\rm exch}(\omega,n_1n_2)&=&\sum_{\mu_c\mu_{n_1}\mu_{n_2}}
I_{vcn_1n_2}(\omega)I_{n_1n_2cv}(\omega-\Delta_{vc})\,,\\
F^{\rm cr}_{\rm dir}(\omega,n_1n_2)&=&\sum_{\mu_c\mu_{n_1}\mu_{n_2}}
I_{cn_2n_1v}(\omega)I_{n_1vcn_2}(\omega)\,,\\
F^{\rm cr}_{\rm exch}(\omega,n_1n_2)&=&\sum_{\mu_c\mu_{n_1}\mu_{n_2}}
I_{vn_2n_1v}(\omega)I_{n_1ccn_2}(\omega-\Delta_{vc})\,,
\ee
$u=(1-i0)$, and the prime on the sum indicates that some terms are excluded
from the summation. First of all, we omit the 
reducible contribution, i.e. the terms for which the intermediate
two-electron energy $\eps_{n_1}+\eps_{n_2}$ equals  the energy of the
initial two-electron state $\eps_v+\eps_c$. Those are:
$(\eps_{n_1}\eps_{n_2})=(\eps_c\eps_v)$ and $(\eps_v\eps_c)$. In addition, we
exclude also the infrared-divergent terms (see
\cite{ShabaevFokeeva94,yerokhin01pra2p} for details), namely those with
$(\eps_{n_1}\eps_{n_2})=(\eps_c\eps_v)$ in the direct crossed part and with
$(\eps_{n_1}\eps_{n_2})=(\eps_c\eps_c)$ and $(\eps_v\eps_v)$ in the exchange
crossed part. These terms should be considered together with the reducible
contribution. Their sum can be shown to be infrared finite. We employ the
notations $\Delta E^{\rm 2el}_{\rm "ir"}$ and $\Delta E^{\rm 2el}_{\rm
"red"}$ in order to emphasize that the
corresponding terms are not "pure" irreducible and reducible
contributions.

We mention that the case under consideration differs from the cases 
of the $2s$ or $2p_{1/2}$ valence electrons considered previously in 
\cite{yerokhin01pra2p}
 by the fact that for the $2p_{3/2}$ Dirac state there is no
adjoining state separated only by the finite-nuclear-size effect. Consequently,
there is no need to exclude any further terms from the crossed contribution, 
as we had to proceed in Ref. \cite{yerokhin01pra2p}
in the case of the $2s$ and the $2p_{1/2}$ valence electron.

Finally, we note the "reducible" contribution 
\be\label{reducile} \Delta E^{\rm 2el}_{\rm
  "red"}&=&\frac{i}{4\pi}\int_{-\infty}^{\infty}d\omega\,
\frac{1}{(\omega+i0)^2}\,\Bigl[2F^{\rm cr}_{\rm exch}(-\omega+\Delta_{vc},cc)
\nonumber \\ &+&
2F^{\rm cr}_{\rm exch}(-\omega,vv) \nonumber \\ &-&
F^{\rm lad}_{\rm exch}(\omega+\Delta_{vc},cv)
-F^{\rm lad}_{\rm exch}(-\omega+\Delta_{vc},cv)
\nonumber \\
&-&F^{\rm lad}_{\rm dir}(\omega-\Delta_{vc},vc)
-F^{\rm lad}_{\rm dir}(-\omega-\Delta_{vc},vc)
\nonumber \\ &-&F^{\rm lad}_{\rm exch}(\omega,vc)-
F^{\rm lad}_{\rm exch}(-\omega,vc)\Bigr]
\,.
\ee

%%%%%%%%%%%%%%%%%%%%%%%%%%%%%%%%%%%%%%%%%%%%%%%%%%%%%%%%%%%
%%%%%%%%%%%%%%%%%%%%%%%%%%%%%%%%%%%%%%%%%%%%%%%%%%%%%%%%%%%
%%%%%%%%%%%%%%%%%%%%%%%%%%%%%%%%%%%%%%%%%%%%%%%%%%%%%%%%%%%
\section{Numerical evaluation}

The three-electron contribution to the energy of $(1s)^22s$,
$(1s)^22p_{1/2}$, and $(1s)^22p_{3/2}$ levels of Li-like ions has been calculated
in our recent investigation \cite{sysak02}. This evaluation appears as
 relatively
simple since the corresponding expressions (\ref{3elirr}), (\ref{3elred})
contain at most one summation over the Dirac spectrum and no integrations
over the virtual-photon energy. Thus we focus here on the calculation
of the two-electron contribution.

The summation over magnetic substates in Eqs.
(\ref{eladdir})-(\ref{ecrexch}), (\ref{reducile}) was performed by means of
standard techniques. The resulting expressions can be found in
\cite{yerokhin01pra2p}. As an independent check we employed also the direct
numerical summation of Clebsch-Gordan coefficients.

To calculate infinite summations over the spectrum of the Dirac equation in
Eqs. (\ref{eladdir})-(\ref{ecrexch}), we employed the method of the B-spline
basis set for the Dirac equation \cite{johnson88}. Typical basis sets
contained 50 positive and 50 negative-energy eigenstates for each value of
the angular-momentum quantum number $\kappa$. The finite size of the nucleus
has been taken into account employing the homogeneously-charged sphere model for
the nuclear-charge distribution. The values of the rms radii used in this
work are the same as in \cite{yerokhin01pra2p}. Infinite summations over
$\kappa$ were truncated typically at $|\kappa|=10$. Partial sums of the expansion
over $|\kappa|$ were fitted to the form
\be
S_{|\kappa|}=a_0+\sum_{n=2}^{N}\frac{a_n}{|\kappa|^n}\,
\ee
using the least squares method. The coefficient $a_0$ yields the extrapolated
value for the sum of the expansion. We found that different fits with $N=$
4-6 yield the same result with an accuracy of at least 5 digits.

\begin{figure*}
\epsfig{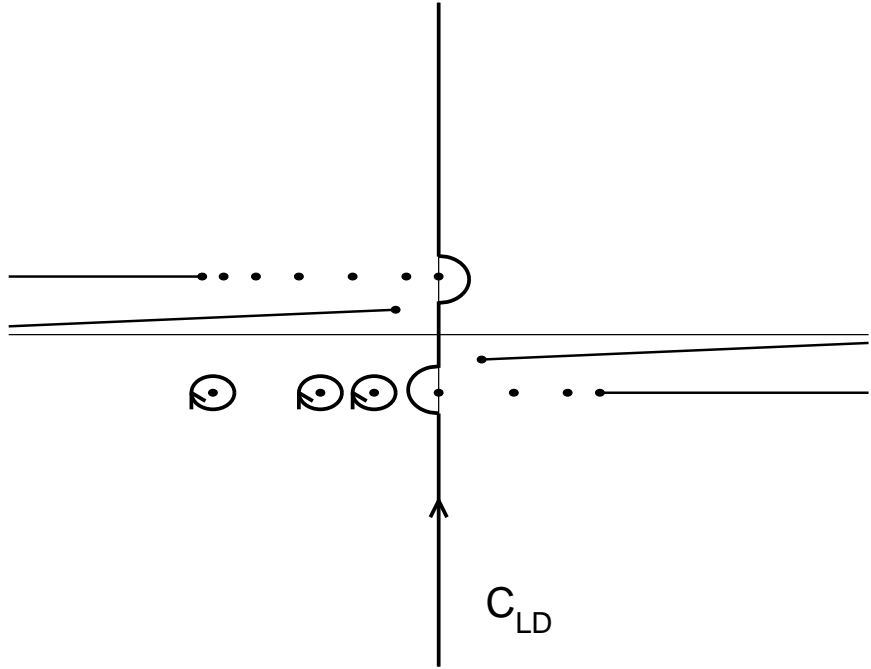}
\caption{\label{lad_dir}
The poles and the branch cuts of the integrand for the direct part of the
ladder contribution, and the integration contour $C_{\rm LD}$.
}
\end{figure*}

\begin{figure*}
\epsfig{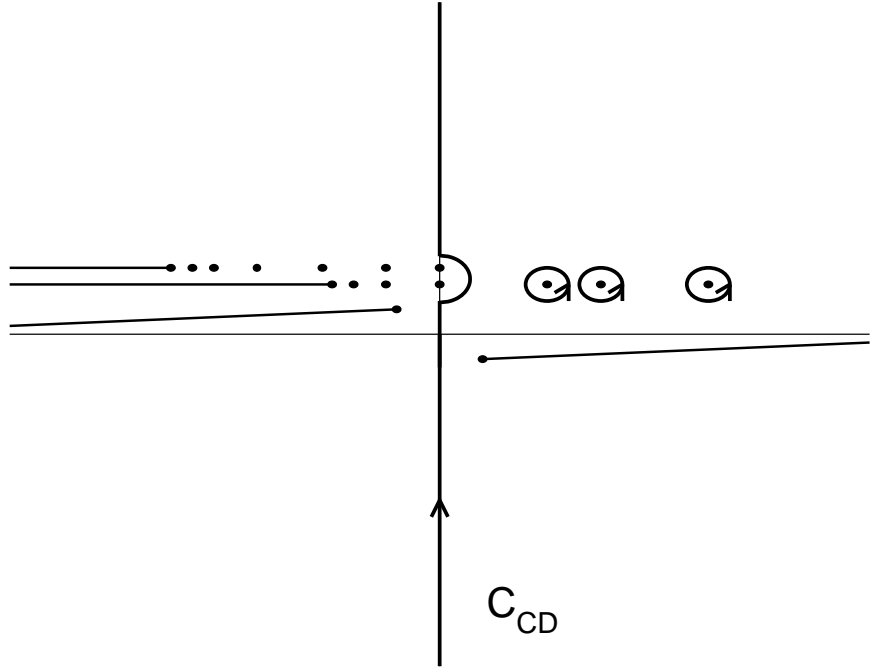}
\caption{\label{cros_dir}
The poles and the branch cuts of the integrand for the direct part of the
crossed contribution, and the integration contour $C_{\rm CD}$.
}
\end{figure*}

\begin{figure*}
\epsfig{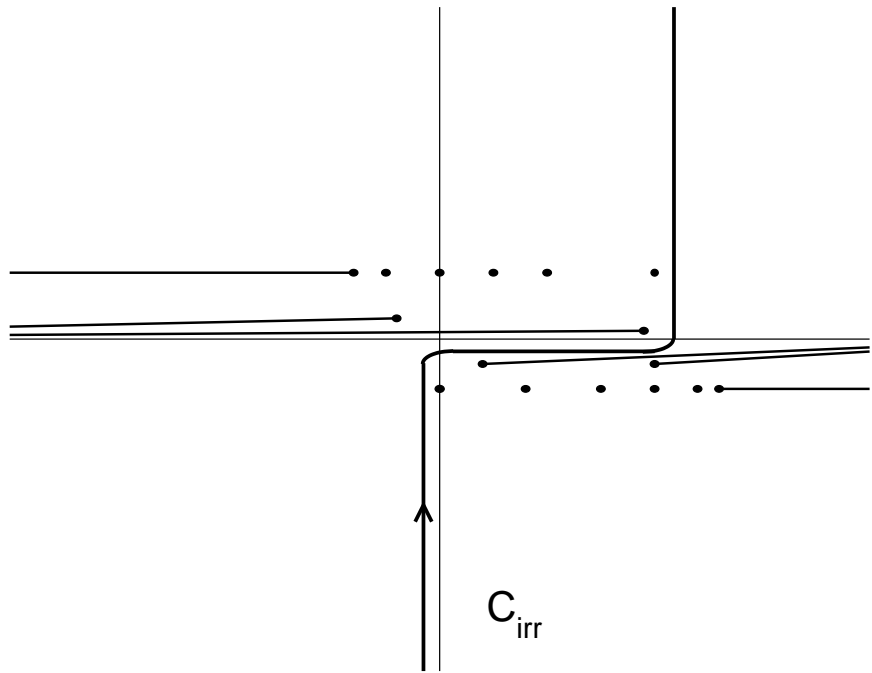}
\caption{\label{lad_exch}
The poles and the branch cuts of the integrand for the exchange part of the
ladder contribution, and the integration contour $C_{\rm irr}$.
}
\end{figure*}

\begin{figure*}
\epsfig{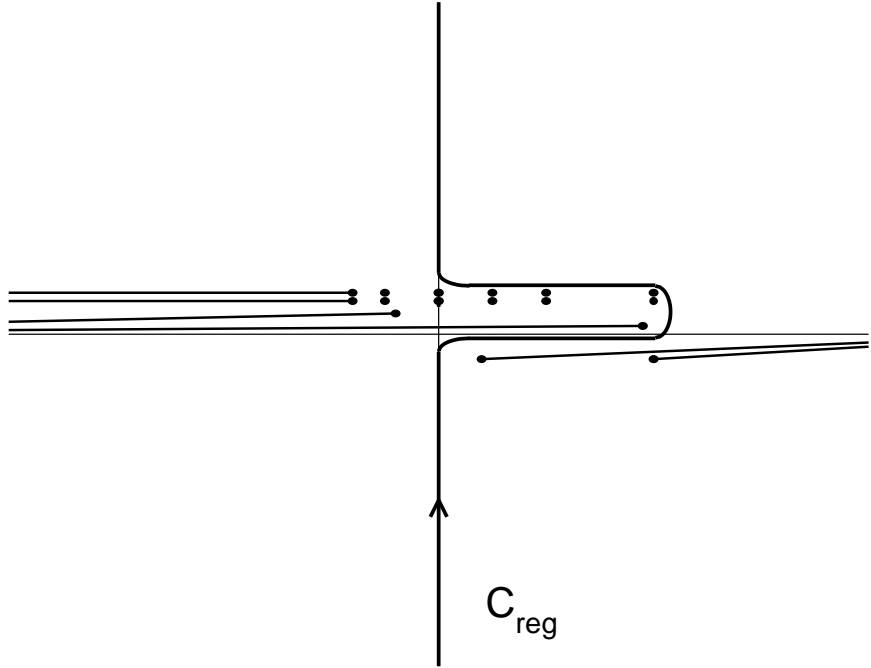}
\caption{\label{cros_exch}
The poles and the branch cuts of the integrand for the exchange part of the
crossed contribution, and the integration contour $C_{\rm reg}$.}
\end{figure*}

The integration over the energy
of the virtual photon $\omega$ in Eqs. (\ref{eladdir})-(\ref{ecrexch})
represents the most difficult part of the calculation. To
avoid strong oscillations for large values of $\omega$, we performed the Wick
rotation of the integration contour. Deforming the contour, one should take
care about the poles and the branch cuts of the integrand. The analytic
structure of the integrand for Eqs. (\ref{eladexch})-(\ref{ecrexch}) is shown
in Figs. \ref{lad_dir}-\ref{cros_exch}. These graphs are very similar to
those for the $2s$- and $2p_{1/2}$-valence electrons in Ref.
\cite{yerokhin01pra2p}. The only difference is that now three Dirac
energy levels occur which are
 more deeply bound than the valence state: $1s$, $2s$, and
$2p_{1/2}$. The terms in Eqs. (\ref{eladexch}) and (\ref{ecrexch}) containing
these states and the valence state as intermediate were treated in a
different way than the remainder, as is discussed below.

For the evaluation of the direct parts of the ladder and crossed
contributions, we perform the Wick rotation of the integration contours
separating the corresponding pole contributions, as shown in Figs.
\ref{lad_dir} and \ref{cros_dir}. In the direct part of the reducible
contribution, we also perform a Wick rotation and then integrate by parts.
This yields the following expression which can be evaluated directly,
\be
\Delta E^{\rm 2el}_{\rm "red",dir}&=& \frac{1}{2}\left[F^{\rm lad}_{\rm
dir}(\Delta_{vc},vc) \right]\pr\nonumber \\ &-&\frac{1}{\pi}\int_0^\infty
d\omega \frac{\omega} {\Delta_{vc}^2+\omega^2}\frac{d}{d\omega}F^{\rm
lad}_{\rm dir}(i\omega,vc)\,,
\ee
where $F\pr(\Delta)=(dF/d\omega)_{\omega=\Delta}$.

Let us now turn to the exchange contribution. As one can see from Figs.
\ref{lad_exch} or \ref{cros_exch}, in this case the integration contour is
squeezed between two branch cuts of the photon propagators on the interval
$[0,\Delta_{vc}]$. Therefore, the standard Wick rotation of the contour is
not possible. It is convenient to divide the contributions of Eqs.
(\ref{eladexch}) and (\ref{ecrexch}) into two parts. The first one
accounts for the poles of the integrand  on the interval
$[0,\Delta_{vc}]$ and is referred to as the {\it irregular} part.
The remainder is denoted as the {\it
regular} part. This contribution does not possess any poles close to the
squeezed part of the contour, which simplifies its numerical evaluation. 
However, it turns out as is the most time-consuming part of the
calculation. One of the integration contours $C_{\rm reg}$ used for the evaluation of 
the regular part is depicted in Fig. \ref{cros_exch}.

The evaluation of the irregular part is less time consuming, but its
structure is more difficult. In this case we need to take care of single and
double poles of the integrand that are located close to the integration
contour. The potential occurrences of one or two single poles and one double pole
within the
interval $[0,\Delta_{vc}]$ were treated by means of the following identities:
\be
\int_{\omega_1}^{\omega_2}d\omega \frac{f(\omega)}{x_0-\omega \pm i0}&=&
P\int_{\omega_1}^{\omega_2}d\omega \frac{f(\omega)}{x_0-\omega}\mp i\pi
f(x_0)\,,\\
\int_{\omega_1}^{\omega_2}d\omega \frac{f(\omega)}{(x_0-\omega \pm i0)^2}
&=&\pm i\pi f\pr(x_0)+\frac{f(\omega_2)}{x_0-\omega_2}\nonumber \\ 
&&-\frac{f(\omega_1)}{x_0-\omega_1}
-P\int_{\omega_1}^{\omega_2}d\omega \frac{f\pr (\omega)}{x_0-\omega}\,,\\
\int_{\omega_1}^{\omega_2}d\omega \frac{f(\omega)} {(x_0-\omega \pm
i0)(x_1-\omega \pm i0)}&=& \frac{1}{x_1-x_0}\Biggl[
P\int_{\omega_1}^{\omega_2}d\omega \frac{f(\omega)}{x_0-\omega}\nonumber\\
&&-P\int_{\omega_1}^{\omega_2}d\omega \frac{f(\omega)}{x_1-\omega}
\mp  i\pi f(x_0) \pm i\pi f(x_1)\Biggr]\,,\label{twopoles}
\ee
where $P$ indicates the principal value of the integral. In Eq.
(\ref{twopoles}) the choice of the sign before $i\pi f(x_0)$ and  $i\pi
f(x_1)$ is determined by the sign of the infinitesimal addition  
$\pm i0$ in the first and the second denominator, respectively. 
For the numerical
evaluation of the irregular contribution we employed the integration contour
$C_{\rm irr}$ shown in Fig. \ref{lad_exch}. It consists of 3 parts:
$[-i\infty-\epsilon,-\epsilon]$, $[-\epsilon, \Delta_{vc}+\epsilon]$, and
$[\Delta_{vc}+\epsilon, \Delta_{vc}+\epsilon+i\infty]$. A
small positive constant $\epsilon$ was introduced
in order to facilitate the numerical
evaluation of the principal-value integrals.

After integration by parts, the exchange contribution of the reducible part
can be written as
\be\label{redexch}
\Delta E^{\rm 2el}_{\rm "red",exch}
%\nonumber \\
&=&-\frac{1}{2}
\Bigl[F^{\rm cr}_{\rm exch}(\Delta_vc,cc)+F^{\rm cr}_{\rm exch}(0,vv)\Bigr]\pr
\nonumber \\ &&+
\frac{1}{2\pi i}P\int_{-\infty}^\infty \frac{d\omega}{\omega}
\frac{d}{d\omega}\Bigl[F^{\rm cr}_{\rm exch}(\Delta_vc+\omega,cc)
\nonumber \\ &&+F^{\rm cr}_{\rm exch}(\omega,vv)- 2 F^{\rm lad}_{\rm exch}(\omega,vc)
\Bigr]\,.
\ee
It is again worth mentioning  that the integral in Eq. (\ref{redexch})
exists only if the sum of all 3 terms in the brackets is considered. For the
each single term, the integral is infrared divergent.

%%%%%%%%%%%%%%%%%%%%%%%%%%%%%%%%%%%%%%%%%%%%%%%%%%%%%%%%%%%%%%%%%%%%%%
%%%%%%%%%%%%%%%%%%%%%%%%%%%%%%%%%%%%%%%%%%%%%%%%%%%%%%%%%%%%%%%%%%%%%%
%%%%%%%%%%%%%%%%%%%%%%%%%%%%%%%%%%%%%%%%%%%%%%%%%%%%%%%%%%%%%%%%%%%%%%
\section{Numerical results and discussion}

The results of our calculations are presented in Table \ref{results}, where
 the direct, the exchange, and the three-electron contribution to the
two-photon exchange correction of the valence $2p_{3/2}$ electron with the
$(1s)^2$ shell are listed separately.
 The evaluation was performed within the Feynman gauge. 
We estimate the numerical uncertainty of our results to be
less than $5\times 10^{-5}$ a.u. For bismuth, our results can be compared
with the calculation by
 Sapirstein and Cheng \cite{sapirstein01}. They report
$-6.529$ and $-6.670$ eV for the two-electron and the three-electron
contribution, respectively. This agrees well with our corresponding
results of $-6.5330$
and $-6.6698$ eV, respectively.

It is interesting to compare the results of the rigorous QED treatment with
approximations evaluations based on relativistic many-body perturbation
theory (MBPT). The difference between the QED and MBPT results can be
conventionally regarded as a "nontrivial" QED contribution. In order to deduce
the two-photon exchange correction within the framework of MBPT, we should
introduce the following changes in our basic formulas: all summations over
intermediate states should be restricted to positive-energy states only, the
calculation should be performed within Coulomb gauge, and the virtual-photon
energy in the photon propagator should be set equal to zero. Within this
approximation, all reducible parts vanish, and the integration over the
energy of the virtual photon can be carried out employing Cauchy's theorem. This
yields zero for the crossed contribution, and finally we are left with the following
expression for the total two-photon exchange correction within the MBPT
approximation: \be \label{MBPT3el} \Delta E_{\rm MBPT}^{\rm
3el}&=&\sum_{PQ}(-1)^{P+Q}\sum_{\eps_n > 0}\!\pr \;\frac{I_{P2P3nQ3}
(0)I_{P1nQ1Q2}(0)}{\eps_{Q1}+\eps_{Q2}-\eps_{P1} -\eps_n}\,,\\
    \label{MBPT2el}
\Delta E_{\rm MBPT}^{\rm 2el}&=&\sum_{\mu_c} \sum_{\eps_{n_1} \eps_{n_2} > 0}\!\!\!\pr\;
\frac{\bigl[I_{cvn_1n_2}(0)-I_{vcn_1n_2}(0)
\bigr]I_{n_1n_2cv}(0)}{\varepsilon_c+\varepsilon_v-
\varepsilon_{n_1}-\varepsilon_{n_2}}\,,
\ee
where the photon propagators should be taken in the Coulomb gauge and the
prime indicates that terms with vanishing denominator should be
omitted. We mention that Eqs. (\ref{MBPT3el}) and (\ref{MBPT2el}) include the
contribution due to the exchange by two Breit photons (the $B\times B$ term).
Strictly speaking, this term is of  higher order than the level of validity of
the Breit approximation, and, therefore, it appears to be inconsistent 
 to include it within the MBPT scheme.

\begin{figure*}
\epsfig{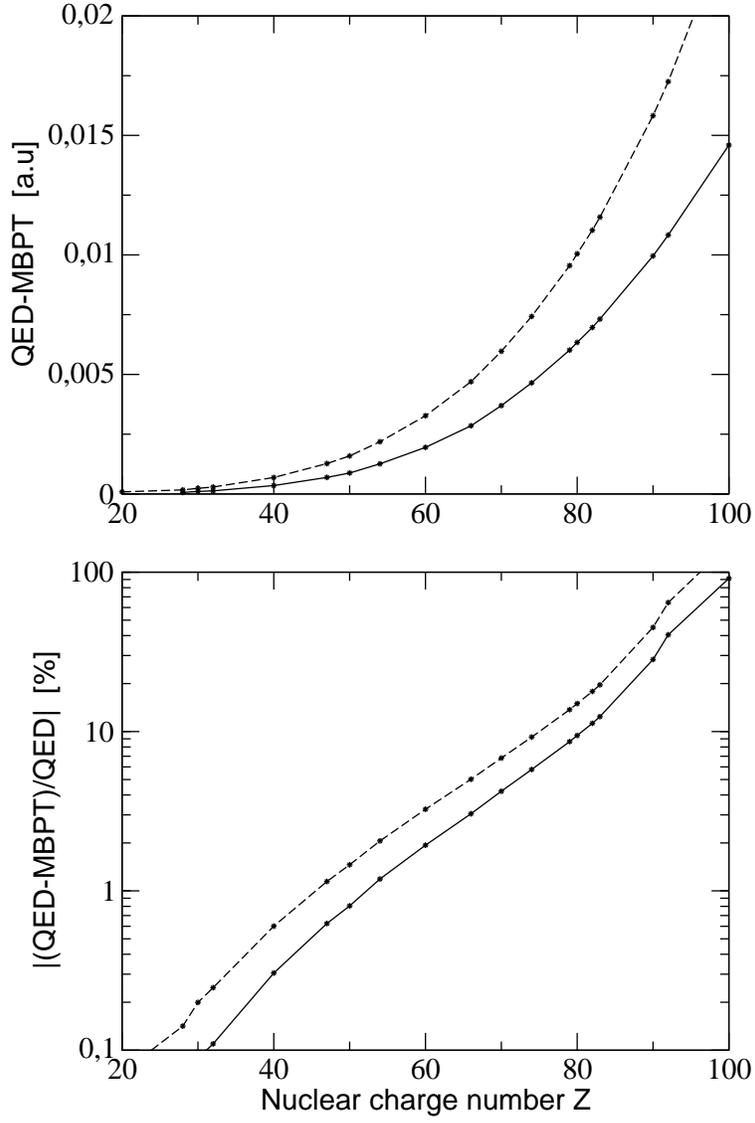} 
\caption{\label{pict_MBPT} The
difference of the QED result for the two-photon exchange correction to the
$2p_{3/2}$-$2s$ transition and the corresponding MBPT results, with the
$B\times B$ term included (solid line) and without this term (dashed line).
The upper graph presents this difference in atomic units, and the lower one
in units of per cent of the total QED contribution.}
\end{figure*}

In Table \ref{tabl_MBPT} and in Fig. \ref{pict_MBPT} we compare the results
of the rigorous QED treatment of the two-photon exchange correction to the
$2p_{3/2}$-$2s$ splitting with the complete MBPT result [Eqs. (\ref{MBPT3el})
and (\ref{MBPT2el})], and the MBPT result dropping the $B\times B$ term.  A
similar analysis for the $2p_{1/2}$-$2s$ splitting has been presented in our previous
investigation \cite{yerokhin01pra2p}. Our results show that in the case under
consideration the nontrivial QED contribution is essentially larger than that
for the $2p_{1/2}$-$2s$ transition. E.g., for uranium it yields $-0.011$
a.u., while the corresponding contribution to the $2p_{1/2}$-$2s$ splitting
is two times smaller, of about $-0.006$ a.u. Moreover, we see that in our
case the total correction changes its sign in the region between $Z=92$ and
100. As a result, the MBPT result becomes incorrect by more than 50\% at very
high values of $Z$. A further conclusion that can be drawn from our comparison
is that for the $2p_{3/2}$-$2s$ splitting the $B\times B$ term is of the same
sign and  magnitude as the nontrivial QED contribution. Thus, its inclusion
improves the agreement between the MBPT and the QED result. This
situation is contrary to the one for the $2p_{1/2}$-$2s$ splitting, where the
$B\times B$ term turns out to be of the same order of magnitude, but of different sign
than the nontrivial QED contribution.

To summarize this investigation we presented a rigorous QED evaluation
of the two-photon exchange correction for the $(1s)^22p_{3/2}$ state of
Li-like ions. Combining these results with the data for the $(1s)^22s$ state
from our previous study \cite{yerokhin01pra2p}, we obtained the two-photon
exchange correction for the $2p_{3/2}$-$2s$ splitting. This is an important
step towards the final goal consisting in the evaluation of all
two-electron second-order QED corrections to the $2p_{3/2}$-$2s$ 
transition energy for the Li isoelectronic sequence.

\section{Acknowledgments}
This work was supported by the Russian Foundation for Basic Research (Grant
No. 01-02-17248), by the Russian Ministry of Education 
(Grant No. E02-3.1-49), and by the program "Russian Universities" (Grant No.
UR.01.01.072). The work of A.N.A. and V.M.S. was supported by the joint grant 
of the Russian Ministry of Education and the Administration of Saint
Petersburg (Grant No. PD02-1.2-79). 
V.A.Y. acknowledges the support of 
the foundation "Dynasty" and
the International Center for Fundamental Physics.
The work of V.M.S. was supported by 
the Alexander von Humboldt Stiftung. 
We also acknowledge
support from BMBF, DFG, DAAD, and GSI.

%%%%%%%%%%%%%%%%%%

\newpage
\begin{table}
\caption{\label{results} Various contributions to the two-photon exchange
correction for the $(1s)^22p_{3/2}$ state of Li-like ions, in atomic units.
The subscripts "dir" and "exch" label the direct and the exchange parts,
respectively; the superscripts "2el" and "3el" refer to the two-electron and
the three-electron contributions, respectively.}
\begin{ruledtabular}
\begin{tabular}{ccccc}
Z&$\Delta E^{\rm 2el}_{\rm dir}$&$\Delta E^{\rm 2el}_{\rm exch}$&
$\Delta E^{\rm 3el}$&Total \\ \hline
 20 &   $\ \ $0.03876 &    0.03902 &   $-$0.45509 &   $-$0.37731 \\
 28 &   $-$0.10511 &    0.03760 &   $-$0.31608 &   $-$0.38359 \\
 30 &   $-$0.12453 &    0.03715 &   $-$0.29807 &   $-$0.38545 \\
 32 &   $-$0.14058 &    0.03673 &   $-$0.28363 &   $-$0.38748 \\
 40 &   $-$0.18304 &    0.03465 &   $-$0.24844 &   $-$0.39683 \\
 47 &   $-$0.20486 &    0.03257 &   $-$0.23449 &   $-$0.40677 \\
 50 &   $-$0.21185 &    0.03156 &   $-$0.23126 &   $-$0.41156 \\
 54 &   $-$0.21978 &    0.03019 &   $-$0.22878 &   $-$0.41836 \\
 60 &   $-$0.22960 &    0.02795 &   $-$0.22801 &   $-$0.42967 \\
 66 &   $-$0.23789 &    0.02560 &   $-$0.22991 &   $-$0.44220 \\
 70 &   $-$0.24292 &    0.02392 &   $-$0.23233 &   $-$0.45133 \\
 74 &   $-$0.24772 &    0.02223 &   $-$0.23553 &   $-$0.46102 \\
 79 &   $-$0.25360 &    0.02001 &   $-$0.24047 &   $-$0.47406 \\
 80 &   $-$0.25477 &    0.01958 &   $-$0.24158 &   $-$0.47677 \\
 82 &   $-$0.25712 &    0.01867 &   $-$0.24390 &   $-$0.48236 \\
 83 &   $-$0.25831 &    0.01822 &   $-$0.24511 &   $-$0.48519 \\
 90 &   $-$0.26682 &    0.01499 &   $-$0.25454 &   $-$0.50636 \\
 92 &   $-$0.26936 &    0.01406 &   $-$0.25752 &   $-$0.51281 \\
100 &   $-$0.28022 &    0.01030 &   $-$0.27070 &   $-$0.54061 \\
\end{tabular}
\end{ruledtabular}
\end{table}

\begin{table}
\caption{\label{tabl_MBPT} Comparison of the rigorous QED treatment of the
two-photon exchange correction to the $2p_{3/2}$-$2s$ splitting in Li-like
ions with the approximate MBPT treatment, in atomic units. ${B\times B}$
denotes the term corresponding to the exchange by two Breit photons. }
\begin{ruledtabular}
\begin{tabular}{cccc}
Z&QED&MBPT&MBPT$-$(${B\times B}$) \\ \hline
 20&$-$0.11912&$-$0.11917&$-$0.11920\\
 28&$-$0.11778&$-$0.11784&$-$0.11794\\
 30&$-$0.11731&$-$0.11741&$-$0.11754\\
 32&$-$0.11681&$-$0.11693&$-$0.11709\\
 40&$-$0.11414&$-$0.11449&$-$0.11483\\
 47&$-$0.11078&$-$0.11147&$-$0.11205\\
 50&$-$0.10897&$-$0.10985&$-$0.11056\\
 54&$-$0.10606&$-$0.10732&$-$0.10825\\
 60&$-$0.10064&$-$0.10258&$-$0.10391\\
 66&$-$0.09355&$-$0.09640&$-$0.09824\\
 70&$-$0.08756&$-$0.09125&$-$0.09352\\
 74&$-$0.08044&$-$0.08508&$-$0.08786\\
 79&$-$0.06960&$-$0.07562&$-$0.07915\\
 80&$-$0.06711&$-$0.07345&$-$0.07715\\
 82&$-$0.06183&$-$0.06879&$-$0.07286\\
 83&$-$0.05898&$-$0.06630&$-$0.07056\\
 90&$-$0.03514&$-$0.04509&$-$0.05096\\
 92&$-$0.02676&$-$0.03759&$-$0.04401\\
100&$\ \ $0.01597&$\ \ $0.00138&$-$0.00781\\
\end{tabular}
\end{ruledtabular}
\end{table}

\newpage

\end{document}